\documentclass[aps,prb,reprint,superscriptaddress,twocolumn]{revtex4-1}

\usepackage{amssymb}
\usepackage{amsmath}
\usepackage{graphicx}
\usepackage{epstopdf}
\usepackage{natbib}
\usepackage{bm}
\usepackage{overpic}
\usepackage[absolute,overlay]{textpos}
\usepackage{ragged2e}
\usepackage[export]{adjustbox}

\newcommand*{\citen}[1]{%
	\begingroup
	\romannumeral-`\x 
	\setcitestyle{numbers}%
	\cite{#1}%
	\endgroup}
\begin{document}


\title{Temperature Dependence of Viscosity in Normal Fluid $^3$He Below 800\,mK Determined by a Micro-electro-mechanical Oscillator}


\author{M. Gonz\'alez}
\email{mgonzalez@phys.ufl.edu}
\altaffiliation{Now at: Aramco Services Company: Aramco Research Center---Houston, 16300 Park Row, Houston, TX 77084, USA }
\affiliation{Department of Physics, University of Florida, Gainesville, FL 32611-8440, USA}
\author{W. G. Jiang}
\affiliation{Department of Physics, University of Florida, Gainesville, FL 32611-8440, USA}
\author{P. Zheng}
\affiliation{Department of Physics, University of Florida, Gainesville, FL 32611-8440, USA}
\author{C. S. Barquist}
\affiliation{Department of Physics, University of Florida, Gainesville, FL 32611-8440, USA}
\author{H. B. Chan}
\affiliation{Department of Physics, Hong Kong University of Science and Technology, Clear Water Bay, Hong Kong, China}
\author{Y. Lee}
\email{yoonslee@phys.ufl.edu}
\affiliation{Department of Physics, University of Florida, Gainesville, FL 32611-8440, USA}



\date{\today}

\begin{abstract}
		A micro-electro-mechanical system vibrating in its shear mode was used to study the viscosity of normal liquid $^3$He from 20\,mK to 770\,mK at 3\,bar, 21\,bar, and 29\,bar. The damping coefficient of the oscillator was determined by frequency sweeps through its resonance at each temperature. Using a slide film damping model, the viscosity of the fluid was obtained. Our viscosity values are compared with previous measurements and with calculated values from Fermi liquid theory. The crossover from the classical to the Fermi liquid regime is manifest in the temperature dependence of viscosity. In the Fermi liquid regime, the temperature dependence of viscosity changes from $T^{-1}$ to $T^{-2}$ on cooling, indicating a transition from the Stokes flow to the Couette flow regime. 
\end{abstract}


\maketitle

\section{Introduction} 
\label{introduction}

Liquid $^3$He has been one of the most important systems on which the foundations of Fermi liquid theory is studied. As a highly pure Fermi system, the low temperature behavior of its transport properties is tied to the strong temperature dependence in the inelastic quasiparticle scattering time \cite{Landau1957,Landau1959}. In the fully degenerate Fermi liquid regime, approximate solutions to the collision integral in the quasiparticle transport equation were first proposed by Abrikosov and Khalatnikov \cite{Abrikosov1959}, and further extended by Hone\cite{Hone1962}. Since then, it has been widely established that the leading term in the viscous relaxation time has a temperature dependence given by $1/\tau\propto T^{-2}$. Extensions to these models to account for higher temperature corrections were later proposed by Emery \cite{Emery1968}, Rice \cite{Rice1967a,Rice1967b}, and others\cite{Pethick1969-2,Ron1970}. 

The viscosity of liquid $^3$He has been experimentally studied using various measurement techniques, from the zero sound attenuation \cite{Abel1966,Pethick1969} to the damping on a moving object immersed in the liquid, such as magnetic vibrating wires \cite{Black1971} and torsional pendulum oscillators \cite{Andronikashvili1946, Betts1963}. These mechanical oscillators have proven instrumental in accurately determining the transport coefficients of the normal liquid $^3$He system. However, while an extended body of studies on its viscous properties can be found in the literature, to date there has not been one where a single viscometer is used to cover a wide temperature and pressure range, and where the crossover between the classical to quantum fluid behavior, \textit{i.e.} the onset of the Fermi liquid behavior, is clearly observed. 

Recently, miniature piezoelectric tuning fork oscillators have become a very valuable tool in the study of liquid helium\cite{Karrai1995, Blaauwgeers2007}. Additionally, other miniature oscillators based on micro and nano-electro-mechanical systems (MEMS and NEMS) have been recently developed by various groups\cite{Gonzalez2013, Collin2010, Collin2011,Rojas2015}. These devices allow the custom engineering of mechanical structures capable of systematically probing quantum fluids at length scales determined by either their transport properties, topological structures such as quantum vortices, or, in the case of the different phases of superfluid $^3$He, their Cooper pair coherence length. For example, there has been recent interest in the properties of both normal liquid $^3$He and superfluid $^3$He confined in a quasi-two-dimensional film\cite{Vorontsov2007, Sharma2011}. 

The MEMS oscillator used in this work consists of a plate suspended above the substrate by four serpentine springs. The device is actuated by the electrostatic interaction between interdigitated electrodes \cite{Gonzalez2013}. The gap between the top oscillating plate and the substrate was designed to have a fixed thickness. The MEMS device is functional without a magnetic field. The unique yet simple geometry of the oscillator ensures that all the electro-mechanical parameters can be obtained analytically and the damping, experienced when oscillating in a fluid, can be fully modeled. The gap between the movable plate and the substrate facilitates the investigation of the entrained liquid film, and makes it possible to examine the behavior of the fluid in a quasi-two-dimensional environment. The large plate size and its one-dimensional movement induce a uniform velocity profile of the moving object preventing additional complications in the analysis often found in other types of oscillators.

The data presented here are extracted from sweeping the excitation frequency through the resonance of the MEMS oscillator from 20\,mK to 770\,mK at 3\,bar, 21\,bar, and 29\,bar. The data are analyzed to give the leading term of the temperature dependent viscosity of liquid $^3$He and then compared with the viscosity values measured in other works as well as the theoretical predictions.
\begin{figure}[]
	\centering
	\includegraphics[width=\linewidth]{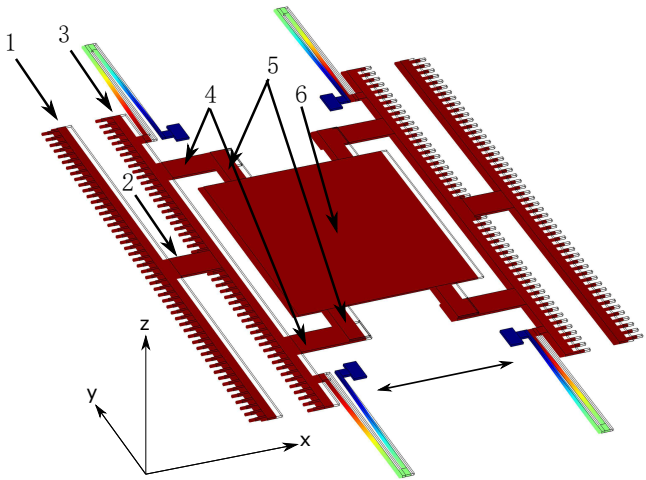}
	\vspace{-30pt}
	\caption{\label{CADfig} 
		A CAD image of the movable part on the MEMS oscillator. The oscillating direction of the plate in the shear mode is marked by the double arrow in the middle. The movable part is composed of six structures labeled on the figure. Structure six is called the oscillating plate and has the largest surface area. Colors show the amplitudes of displacement of various structures. The red movable parts are oscillating with the largest amplitude. The four blue squares are fixed on the substrate, representing the anchors of the four serpentine springs.}
	\raggedbottom
\end{figure}

\section{Analytical description of the MEMS device}
\label{mems}
\subsection{Device characteristics and slid film damping model}
\label{modeldevice}
The movable part of the oscillator can be divided into six types of structures, as shown in Fig.~\ref{CADfig}. Each one contributes differently to the damping due to their difference in size, aspect ratio with respect to the oscillation direction, and different gap distance from the substrate. A film of liquid $^3$He is formed in the gap beneath these structures.
Additional information about the device and the measurement scheme can be found in reference\,~[\citen{Gonzalez2013}, \citen{Barquist2014}]. 
In vacuum, at least four well-separated eigenfrequencies were identified in the oscillator through COMSOL multiphysics simulations. Each one corresponds to a different oscillation mode of the movable plate: the trampoline mode with vertical motions, two pivot modes with rotational oscillations along the $x$-axis and the $y$-axis, respectively, and the shear mode with horizontal oscillations along the $x$-axis as shown in Fig.~\ref{CADfig}. When the device is immersed in liquid, only the shear mode can be observed due to the high damping in all the other modes.

\begin{table*}[]
	\caption{\label{GapArea} Values of gap sizes and areas of various structures of the movable part. The labels from 1 to 6 are the same as those in Fig.~\ref{CADfig}}
	\begin{ruledtabular}
		\begin{tabular}{cccc}
			\raggedbottom
			Structure $i$ & $\tilde{d}_i$ ($\mu$m) & $d_i$ ($\mu$m) &$A_i$ ($\mu$m$^2$) \\
			\hline
			S1 & 2 & 0.90 & $2\times398\times14$=11144 \\
			S2 & 2 & 1.39 & $2\times39\times24$=1872 \\
			S3 & 2 & 0.90 & $2\times400\times14$=11200 \\
			S4 & 2 & 1.60 & $4\times67\times22$=5896 \\
			S5 & 2.75 & 1.27 & $4\times20\times13$=1040 \\
			S6 & 0.75 & 0.72 & $178\times178$=31684 \\
		\end{tabular}
	\end{ruledtabular}
\end{table*}

The steady flow of an incompressible fluid can be described by the Navier-Stokes equation. It describes the most general case of a steady flow of an incompressible fluid \cite{LandauFM}:
\begin{equation}
\rho\left[\frac{\partial \bm{v}}{\partial t}+(\bm{v}\cdot \nabla)\bm{v}\right]=\bm{F}-\nabla p+\eta \nabla^2\bm{v},
\label{NSeqn}
\end{equation}
where $\bm{v}$ is the velocity field, $\rho$ is the fluid density, $\bm{F}$ is the external force, $p$ is the pressure of the liquid, and $\eta$ is the dynamic viscosity. In our experiment, the plate is submerged in a fluid and in relative tangential motion to a fixed substrate. The direction of the one-dimensional oscillation of the plate is defined as the $x$-direction. In the absence of $\bm{F}$ and $\nabla p$, Eq.~(\ref{NSeqn}) reduces to
\begin{equation}
\frac{\partial v_x}{\partial t}+v_x\frac{\partial v_x}{\partial x}=\frac{\eta}{\rho}\frac{\partial^2v_x}{\partial z^2}.
\label{NSeqn2}
\end{equation}
Here, $z$ is the direction perpendicular to the substrate surface with the origin at the substrate. In the real experimental system, the length scale of the plate is much larger than both the gap size and the amplitude of oscillation. Therefore, the velocity can be taken to be translationally invariant in the $x$-direction, and Eq.~(\ref{NSeqn2}) can be further simplified by eliminating the $\partial v_x/\partial x$ term:
\begin{equation}
\frac{\partial v_x}{\partial t}=\frac{\eta}{\rho}\frac{\partial^2v_x}{\partial z^2}.
\label{NSeqn3}
\end{equation}
Equation~(\ref{NSeqn3}) is a one-dimensional diffusion equation, which can be solved in two different flow regimes: the Couette and the Stokes flow regimes. 


When the gap size, $d$, is much smaller than the viscous penetration depth $\delta=\sqrt{2\eta/\rho\omega}$, referred to as the Couette regime, Eq.~(\ref{NSeqn3}) can be approximated to
\begin{equation}
\frac{\partial^2 v_x}{\partial z^2}=0.
\label{NSCouette}
\end{equation}
If the boundary condition is non-slip, Eq.~(\ref{NSCouette}) has a simple solution:
\begin{equation}
v_x(z)=v_p\frac{z}{d}
\label{vCouette}
\end{equation}
where $v_p$ is the velocity of the oscillating plate. The solution represents a simple linear velocity profile inside the gap. From this solution, the viscous force on the plate can be calculated as $F=A_p\eta v_p/d$, where $A_p$ is the contact surface area of the plate to the liquid. The damping coefficient is then simply $\gamma=A_p\eta/d$.

On the other hand, in the Stokes limit where the condition $\delta \gg d$ is not satisfied, no approximation can be made in Eq.~(\ref{NSeqn3}). By implementing the non-slip boundary conditions again, the flow field is given by \cite{Veijola2001}
\begin{equation}
v_x(z)=v_p\frac{\sinh(qz)}{\sinh(qd)},
\label{vStokes}
\end{equation}
where $q=\sqrt{i\omega/\nu}$ and $\nu=\eta/\rho$. The damping admittance can be calculated in a similar way as above:
\begin{equation}
\bar{\gamma}=\frac{F}{v_p}=A_p\frac{\eta}{v_p}\frac{\partial v_x}{\partial z}|_{z=d}=\frac{A_pq\eta}{\tanh(qd)}.
\label{gammaStokes}
\end{equation}
The real part  of this complex admittance gives the damping coefficient
\begin{equation}
Re(\bar{\gamma})=\frac{\eta A_p}{\delta}\frac{\sinh(2d/\delta)+\sin(2d/\delta)}{\cosh(2d/\delta)-\cos(2d/\delta)},
\label{gammareal}
\end{equation}
and the imaginary part is responsible for the frequency shift due to mass loading
\begin{equation}
Im(\bar{\gamma})=\frac{\eta A_p}{\delta}\frac{\sinh(2d/\delta)-\sin(2d/\delta)}{\cosh(2d/\delta)-\cos(2d/\delta)}.
\label{gammaimag}
\end{equation}

Taking into account all six structures of the movable part of the MEMS oscillator (see Fig.\,\ref{CADfig}), the dependence of damping coefficient on the viscosity can be expressed as
\begin{equation}
\gamma_{tot}\\=\frac{\eta}{\delta}\left[\sum\limits_{i=1}^6 A_i\frac{\sinh(\frac{2d_i}{\delta})+\sin(\frac{2d_i}{\delta})+k_{1i}}{\cosh(\frac{2d_i}{\delta})-\cos(\frac{2d_i}{\delta})+k_{2i}}+A_t\right],
\label{gammatot}
\end{equation}
where $\gamma_{tot}$ is the total damping coefficient on the moving plate, $d_i$ is the effective gap size and $A_i$ is the area of the $i^{th}$ structure of the movable part  in contact with the confined liquid film
or the top/bottom side of the movable part, $k_{1i}$ and $k_{2i}$ are two parameters that arise when first order slip boundary conditions are considered. $A_t=\sum_{i=1}^{6}A_i$ is the total area of the movable part in contact with the bulk fluid and this term in Eq.~(\ref{gammatot}) accounts for the damping contribution from the bulk fluid above the movable part. The effective gap size, $d_i$, accounts for the effect of the finite size of the structures on the MEMS oscillator. It was calculated from the real gap size $\tilde{d}_i$ and the length $l$ of the corresponding structure in the direction of oscillation \cite{Veijola2001}:
\begin{equation}
d_i=\frac{\tilde{d}_i}{1+8.5\frac{\tilde{d}_i}{l}}.
\label{di}
\end{equation}
The $d_i$ and $A_i$ values are taken from the design with an error of 5\%, listed in Table~\ref{GapArea}.

The forms of $k_{1i}$ and $k_{2i}$ are \cite{Veijola2001}
\begin{widetext}
	\begin{eqnarray}
	&k_{1i}=4r[(1+r^2)\cosh(\frac{2d}{\delta})+(1-r^2)\cos(\frac{2d}{\delta})]+6r^2[\sinh(\frac{2d}{\delta})-\sin(\frac{2d}{\delta})],\\
	&k_{2i}=4r[(1+2r^2)\sinh(\frac{2d}{\delta})+(1-2r^2)\sin(\frac{2d}{\delta})]+4r^2[(2+r^2)\cosh(\frac{2d}{\delta})+(2-r^2)\cos(\frac{2d}{\delta})].
	\end{eqnarray}
\end{widetext}
\noindent Here, $r=\zeta/\delta$ is the ratio of slip length to the penetration depth. The slip length is a phenomenological parameter introduced as a correction to the hydrodynamic boundary condition and is of the order of the mean free path \cite{Einzel1990,Einzel1997}. Around 100\,mK, the slip length is $\sim$10\,nm and the penetration depth is $\sim$1\,$\mu$m. Therefore in this experiment $r\sim 10^{-2}\ll1$ and $k_{1i}$ and $k_{2i}$ are at least two orders of magnitude smaller than the other terms in Eq.~(\ref{gammatot}), therefore negligible. In the limit $\delta\ll d$, Eq.~(\ref{gammatot}) reduces to $\gamma=A_p\eta/\delta$, where $A_p=2A_t$ now includes both the top and the bottom sides of the plate.

In this experiment, the thickness $d$ of the gap and the penetration depth $\delta$ are comparable. Therefore, the full expression Eq.~(\ref{gammatot}) should be applied. This then enables us to study the temperature dependence of the viscosity in normal fluid $^3$He by applying Khalatnikov's prediction for the viscosity at low temperatures \cite{Abrikosov1959}:
\begin{equation}
\frac{1}{\eta T^2}=a,
\label{Khalatnikov}
\end{equation}
where $a$ is a function of pressure. One can recast Eq.~(\ref{gammatot}) in a form more suitable for analysis by using $P_1=\sqrt{1/a}$ and $P_2=\sqrt{2/\rho\omega}$:
\begin{equation}
	\gamma_{tot}=\frac{P_1}{P_2 T}\left[\sum\limits_{i=1}^6 A_i\frac{\sinh(\frac{2d_i T}{P_1 P_2})+\sin(\frac{2d_i T}{P_1 P_2})}{\cosh(\frac{2d_i T}{P_1 P_2})-\cos(\frac{2d_i T}{P_1 P_2})}+A_t\right]+P_3.
	\label{gammatotalb}
	\end{equation}
Here $P_3$ is an unknown constant background term. $P_2$ is a constant that can be directly calculated. At a given pressure, one can use the tabulated molar volume of liquid $^3$He to find its density, $\rho$ \cite{Halperin1990}. And $\omega$ can be taken as the averaged resonance frequency over the whole temperature range since the shift is typically only within 3\%. Hence, we get the values of $P_2$ as listed in Table~\ref{vrhop2}.
\begin{table*}[]
	\caption{\label{vrhop2} Values of molar volumes, densities, and $P_2$ at various pressures.}
	\begin{ruledtabular}
		\begin{tabular}{ccccc}
			Pressure (bar) & Molar volume (cm$^3$/mol) & $\rho$ (kg/m$^3$) &$f$ (Hz) &$P_2$ ($\times10^{-4}$(m$^3\cdot$s/kg)$^\frac{1}{2}$) \\
			\hline
			3 & 33.9 & 88.9 & 23103 & 3.93 \\
			21 & 27.5 & 110 & 22999 & 3.56 \\
			29 & 26.3 & 115 & 22958 & 3.48 \\
		\end{tabular}
	\end{ruledtabular}
\end{table*}
\begin{figure}[]
	\centering
	\includegraphics[width=\linewidth]{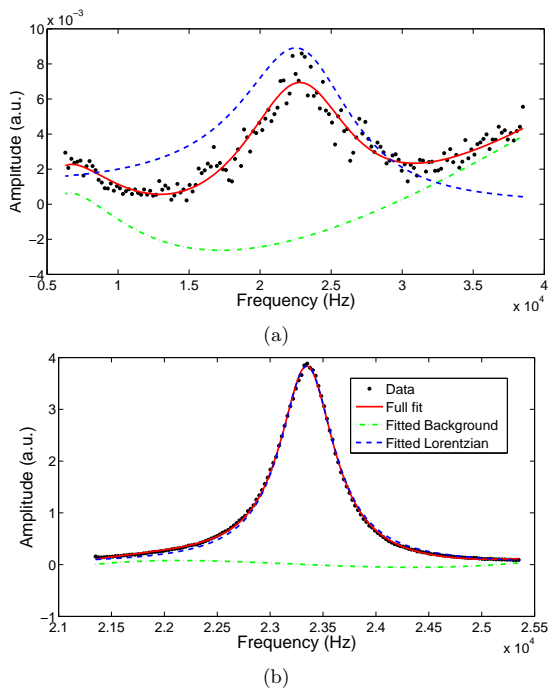}
	\vspace{-50pt}
	\caption{\label{FitCheckH2}  
		(a) Signal amplitude vs frequency at 25\,mK, 21\,bar. The full fit is plotted along with the fitted Lorentzian component and the background. (b) Signal amplitude vs frequency at 758\,mK, 21\,bar. 
	}
\end{figure}
\subsection{Resonance model}
\label{measurescheme}
\begin{figure}[]
	\centering
    \includegraphics[width=\linewidth]{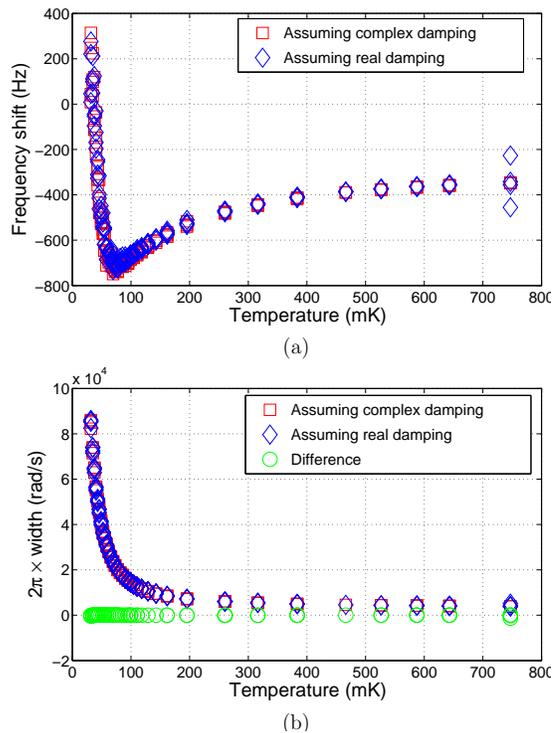}
	\vspace{-40pt}
	\caption{\label{compare}  
		Another way of formulating the fitting model is to introduce a complex damping coefficient $\gamma=m(\alpha_2+i\alpha_1)$, where the imaginary term, $\alpha_1f/2\pi$, explicitly represents the shift in the resonance frequency from the natural frequency ($f_{n}$). Both real and imaginary damping coefficient methods are shown here. (a) Frequency shift vs Temperature from two models at 3\,bar. The frequency shift with a real damping is calculated as $f_0-f_n$. The frequency shift with a complex damping is calculated as $\alpha_1/4\pi$. ($\Box$): include the imaginary part of the damping coefficient. Shift for this curve is $\alpha_1/4\pi$. ($\diamond$): only include the real damping coefficient. Shift for this curve is $f_0-f_n$. (b) Width vs Temperature. ($\Box$): include the imaginary part of the damping coefficient. ($\diamond$): only include the real damping coefficient. ($\bigcirc$): width($\Box$)-width($\diamond$) }
\end{figure}
\begin{figure*}[]
	\centering
	\includegraphics[width=0.97\textwidth]{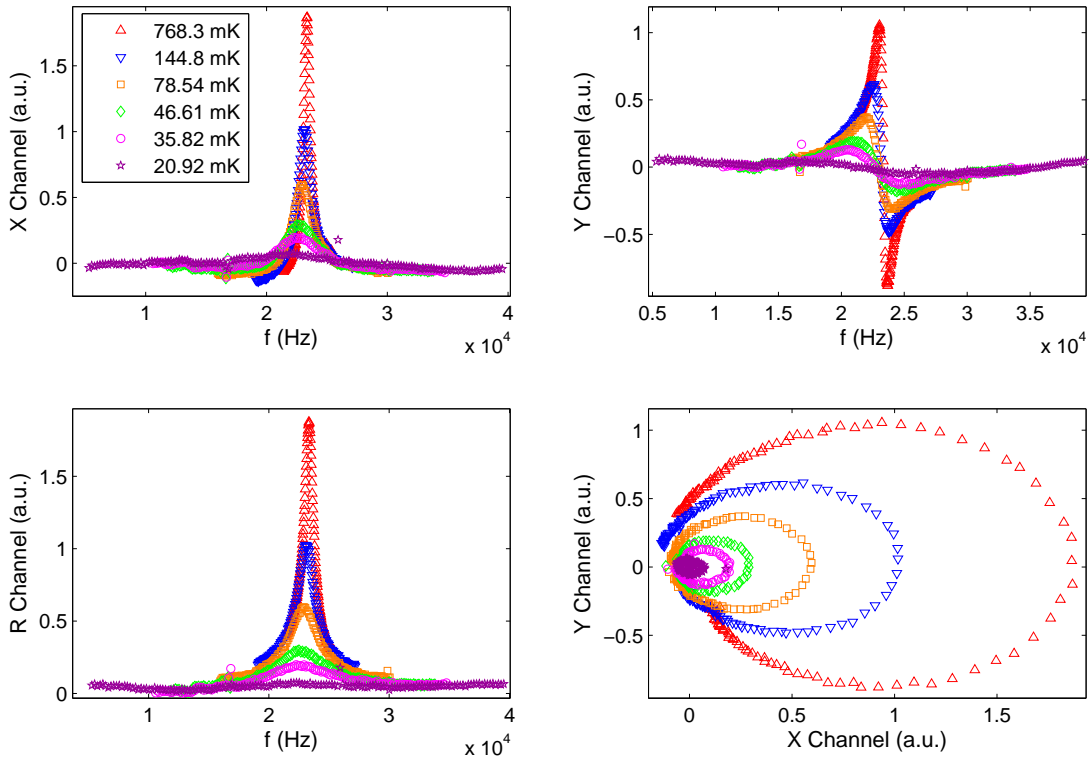}
	\caption{\label{Spectra} 
		(a) to (c) The frequency spectra measured at different temperatures at 29\,bar. Both the resonance frequency shift and the width increase can be observed. (d) The Nyquist diagram: a plot of the Y-channel signal vs the X-channel signal. It shows a frequency dependent ellipse consistent with the model and can be utilized as a guide for phase adjustment.}
\end{figure*}
The pressure of the liquid was maintained at three different values: 3\,bar, 21\,bar, and 29\,bar. Frequency sweeps near 23\,kHz were carried out to obtain a resonance Lorentzian peak proportional to the amplitude of  displacement of the moving plate. The MEMS device is driven by a periodic external force, $F_e=F_0e^{-i\omega t}$. The experiment was performed in the hydrodynamic limit, $\omega\tau\ll1$, where $\omega$ is the driving frequency and $\tau$ is the quasi-particle relaxation time. The oscillation of the center plate of the device can be described by
\begin{equation}
	m\ddot{x}(t)+\gamma\dot{x}(t)+kx(t)=F_e .
	\label{MEMSmotion}
\end{equation}
Here, $m$ is the total mass of the movable part of the MEMS device, $\gamma$ is the damping coefficient, $k$ is the total spring constant of the system, and $x(t)$ is the displacement of the center plate measured from its rest position. The solution to Eq.~(\ref{MEMSmotion}) gives the magnitude of displacement as
\begin{equation}
	\vert x\vert=\frac{F_0}{m}\frac{1}{4\pi^2\sqrt{(f_0^2-f^2)^2+w^2f^2}} ,
	\label{dispX}
\end{equation}
where $w=\gamma/2\pi m$. This solution establishes that $\vert x\vert$ is a peaked function of center frequency $f$ and with a width $w$. 

Given the knowledge of $m$ and $w$, one can calculate the damping coefficient $\gamma=2\pi mw$. $w$ is obtained from the fitting of this function. Details about the measurement circuit and the detection scheme can be found in the reference\,~[\citen{Gonzalez2013}, \citen{Barquist2014}]. $\vert x\vert$ (R-channel) is usually measured through two components, the quadrature component (X-channel),
\begin{equation}
	x_x=\frac{F_0}{m}\frac{wf}{4\pi^2((f_0^2-f^2)^2+w^2f^2)} ,
	\label{disperse}
\end{equation}
and the in-phase component (Y-channel), 
\begin{equation}
	x_y=\frac{F_0}{m}\frac{f_0^2-f^2}{4\pi^2((f_0^2-f^2)^2+w^2f^2)} .
	\label{absorb}
\end{equation}
The frequency dependence of $\vert x\vert^2=x_x^2+x_y^2$ is fitted with a background $BG=a_1f+a_0+c_1/f+c_2/f^2$.
This background originates from the measurement circuit.

One can combine $F_0$ and $m$ into one parameter $A=F_0/m$ to obtain the finalized fitting model:
\begin{equation}
	\vert x\vert^2=\frac{A^2}{16\pi^4}\frac{1}{(f_0^2-f^2)^2+(w^2f^2)^2}+BG .
	\label{fitmodel1}
\end{equation}
There are seven fitting parameters in the model: $A$, $f_0$, $w$, $a_1$, $a_0$, $c_1$, and $c_2$. Typical fittings at high and low temperatures are depicted in Fig.~\ref{FitCheckH2}.

At a given temperature, we performed four frequency sweeps. Their fitting results are averaged to give the fitted parameters at this temperature. Due to the sensitivity of the model to its initial input parameters, a regression fitting issue arises where a set of four fitted parameters may have a variance larger than expected. This happens even though all four sweeps have very similar curve shapes. An outlier curve can be spotted from the four fitted backgrounds. The problem is solved by manually fixing the outlier's background to be the average of the others and fit the Lorentzian part of the curve after subtracting the background. Fig.~\ref{compare} shows the obtained fitting parameters for 3\,bar.

%

\section{Measurements and results}
\label{results}

Four sweeps were performed at a given temperature: two sweeps were done with increasing frequency and two with decreasing frequency. Some of the measured frequency spectra are plotted in Fig.~\ref{Spectra}. 
As the temperature is lowered, the viscosity is expected to follow a $1/T^2$ relation, increasing the damping force. As a result, the Lorentzian peak broadens. The R-channel also shows that the resonance frequency decreases. 
As temperature drops, the viscous penetration depth increases, resulting in a larger effective mass of the oscillator and a decrease in the resonance frequency. 
\begin{table*}[]
	\caption{\label{FitP123} Fitted values of $P_1$, $P_3$ and $1/a$ directly obtained from $P_1$. Results from Wheatley \cite{Wheatley1975} are also presented. Note that Wheatley's third $1/a$ value, 0.99\,Poise$\cdot$mK$^2$, was obtained at 30\,bar instead of 29\,bar.}
	\begin{ruledtabular}
		\begin{tabular}{ccccc}
			Pressure & $P_1$  & $P_3$  & $1/a$  & Wheatley's result \\
			(bar) & ($\times 10^{-4}$\,(Pa$\cdot$ s)$^{\frac{1}{2}}$K) & ($\times10^{-6}$ (kg/s)) & (Poise$\cdot$mK$^2$) & (Poise$\cdot$mK$^2$) \\
			\hline
			3 & 4.40 & 1.38 & 1.93 & 1.73 \\
			21 & 3.45 & 1.57 & 1.19 & 1.22 \\
			29 & 3.05 & 1.42 & 0.93 & 0.99 \\
		\end{tabular}
	\end{ruledtabular}
\end{table*}

\begin{figure}[]
	\begin{overpic}[scale=0.335,tics=10]{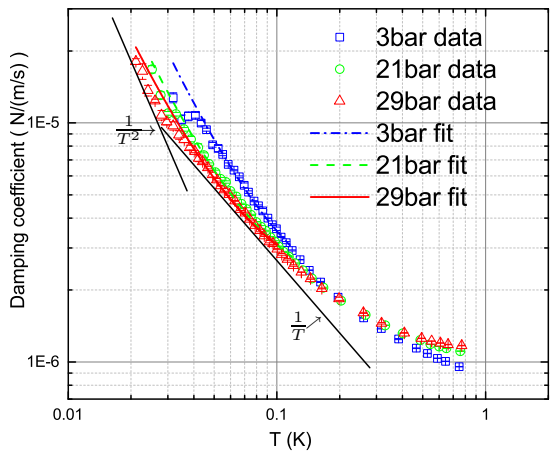}		   
		\put(20,58){\small $\frac{1}{T^2}$}
		\put(24,58){\scriptsize $\rightarrow$}
		\put(52,22){\small $\frac{1}{T}$}
		\put(55,23){\tiny $\nearrow$}
	\end{overpic}
	\caption{\label{1forall} 
		Damping vs Temperature at 3, 21 and 29\,bar. The dashed curves are fitted with the instrumental weight $1/\sigma^2$, where $\sigma$ is the standard deviation of the width based on the four sweeps at each temperature. The solid lines are $1/T$ and $1/T^2$ guidelines. There is a crossing point at 200\,mK. At temperatures higher than this point, higher pressures give larger damping. Below this point, the order in pressure dependence is reversed. The damping values correspond to the Lorentzian widths ranging from $\sim600$\,Hz at the highest temperature to $\sim10000$\,Hz at the lowest.}
\end{figure}

The values of $P_1$ are determined by fitting Eq.~(\ref{gammatotalb}) in the Fermi liquid regime. The fit was done below 100\,mK at all pressures. 
 It was performed with an instrumental weight $\kappa=1/\sigma^2$ where $\sigma$ is the standard deviation of the four damping coefficients obtained at each temperature. 
Our data and the fitted curves are plotted in Fig.~\ref{1forall}.
The fitted values of $P_1$, $P_3$, and $1/a=P_1^2$ are listed in Table~\ref{FitP123}.
\begin{figure}[]
	\includegraphics[width=0.5\textwidth]{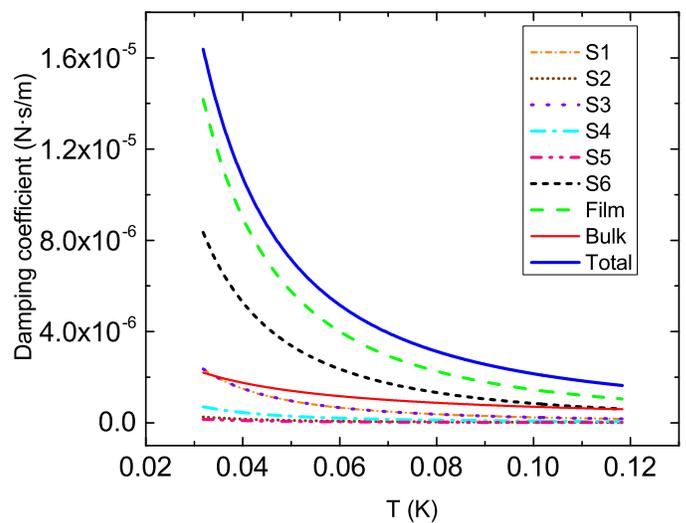}
	\caption{\label{DBP} 
		Damping contribution from the individual structures in the MEMS device as a function of temperature at 3\,bar. These were calculated from Eq.~(\ref{gammatotalb}) using fitted $P_1$ and calculated $P_2$ values. The film contribution is the sum of the terms S1 through S6. It shows that the major damping contribution comes from the film; and that the major contribution to the film damping comes from the moving plate S6. Data for the other two pressures behave very similarly.}
\end{figure}
Our results of $1/a$ are comparable to the theoretical estimation made by Abrikosov and Khalatnikov which is $\sim$1--10\,poise$\cdot$mK$^2$\, and also to previous results found in reference\,~[\citen{Wheatley1975}]. The result from the zero-sound data of Ketterson \textit{et al.} gives a similar value of 1.04\,poise$\cdot$mK$^2$ at 29.3\,bar \cite{Ketterson1975}.


We can compare the damping contributions from different structures of the oscillator. The temperature dependence of the individual damping coefficients are plotted in Fig.~\ref{DBP}. The damping values are calculated using Eq.~(\ref{gammatotalb}) with the fitted $P_1$ and calculated $P_2$ values. The film damping dominates at all pressures. 

In Fig.~\ref{1forall}, the damping coefficient exhibits a clear crossover from $1/T$ to $1/T^2$ behavior as the liquid cools down. Much like the Hagen-Poiseuille to Knudsen crossover observed in nanoholes and aerogel \cite{Savard2009, Takeuchi2012}, this phenomenon is closely related to the length scales in the MEMS-liquid system \cite{Einzel1998}. In the high temperature limit where $\delta\ll d$, $\gamma\propto \eta/\delta\propto\sqrt{\eta}\propto 1/T$; and in the low temperature limit where $\delta\gg d$, $\gamma\propto \eta/d\propto 1/T^2$. Thus, the temperature dependence of the damping coefficient agrees with expectations.

It is interesting to observe that the damping curves for the three pressures intersect at around 200\,mK (see Fig.~\ref{1forall}), indicating a reversal in the pressure dependence of the viscosity. 
This inversion of pressure dependence implies a crossover from classical to quantum fluid. In the classical regime, the viscosity generally increases with pressure. 
While in the Fermi liquid regime, the dominant pressure dependence arises from the velocity, which is the Fermi velocity that decreases with pressure. To our knowledge there is no theoretical representation to fully describe the cross-over behavior.

\section{Summary}
\label{summary}
The damping of a shear actuated MEMS oscillator submerged in liquid $^3$He  was studied. The device moves parallel to the substrate and maintains a constant gap of 0.75~$\mu$m. The parallel plate geometry of the device allows for a full analytical description of its interaction with the fluid through the so-called slide film damping model. The resonance properties were studied through a wide range of temperatures, from 20 to 800~mK, and at three different pressures: 3~bar, 21~bar, and 29~bar. As the liquid is cooled, a crossover in the pressure dependence of the damping occurs around T$=200$~mK. Below this temperature, the damping decreases with pressure as opposed to increasing with pressure. This indicates a transition from a classical to a quantum fluid behavior, where the dominant pressure dependence of the viscosity is determined by the Fermi velocity. To the best of our knowledge, this is the first time this transition is discernibly captured with a single viscometer. 

An extensive analysis was performed to determine the different contributions to the damping from the different structures of the device. Relative to the length scales set by the viscous penetration depth and the gap size of the device, transitions from a Couette to a Stokes flow regime were seen as the viscosity changes from a $1/T$ to $1/T^2$ behavior at lower temperatures. Our findings show the potential of using these devices to test fundamental fluid mechanical problems by exploiting both the strong temperature dependence of the transport properties of liquid $^3$He and the tunability of the characteristic length scales through the engineering of the device geometry. For instance, at temperatures lower than the ones presented in this work, where the mean free path becomes comparable to the gap size, these devices could be used to study the so-called Knudsen regime. Furthermore, in the superfluid state of liquid $^3$He, the devices could be used to study the effects of confining liquid $^3$He to a 2D geometry, where the thickness of the entrained fluid film is comparable to the coherence length of the superfluid phase. This would enable the exploration of novel phase transitions and topological phenomena in unconventional superfluids.

\begin{acknowledgments}
We are very grateful to Erkki Thuneberg and Priya Sharma for the many discussions that helped clear out the theoretical details.

This work is supported by NSF DMR-1205891 (YL).
\end{acknowledgments}


\begin{thebibliography}{33}%
\makeatletter
\providecommand \@ifxundefined [1]{%
 \@ifx{#1\undefined}
}%
\providecommand \@ifnum [1]{%
 \ifnum #1\expandafter \@firstoftwo
 \else \expandafter \@secondoftwo
 \fi
}%
\providecommand \@ifx [1]{%
 \ifx #1\expandafter \@firstoftwo
 \else \expandafter \@secondoftwo
 \fi
}%
\providecommand \natexlab [1]{#1}%
\providecommand \enquote  [1]{``#1''}%
\providecommand \bibnamefont  [1]{#1}%
\providecommand \bibfnamefont [1]{#1}%
\providecommand \citenamefont [1]{#1}%
\providecommand \href@noop [0]{\@secondoftwo}%
\providecommand \href [0]{\begingroup \@sanitize@url \@href}%
\providecommand \@href[1]{\@@startlink{#1}\@@href}%
\providecommand \@@href[1]{\endgroup#1\@@endlink}%
\providecommand \@sanitize@url [0]{\catcode `\\12\catcode `\$12\catcode
  `\&12\catcode `\#12\catcode `\^12\catcode `\_12\catcode `\%12\relax}%
\providecommand \@@startlink[1]{}%
\providecommand \@@endlink[0]{}%
\providecommand \url  [0]{\begingroup\@sanitize@url \@url }%
\providecommand \@url [1]{\endgroup\@href {#1}{\urlprefix }}%
\providecommand \urlprefix  [0]{URL }%
\providecommand \Eprint [0]{\href }%
\providecommand \doibase [0]{http://dx.doi.org/}%
\providecommand \selectlanguage [0]{\@gobble}%
\providecommand \bibinfo  [0]{\@secondoftwo}%
\providecommand \bibfield  [0]{\@secondoftwo}%
\providecommand \translation [1]{[#1]}%
\providecommand \BibitemOpen [0]{}%
\providecommand \bibitemStop [0]{}%
\providecommand \bibitemNoStop [0]{.\EOS\space}%
\providecommand \EOS [0]{\spacefactor3000\relax}%
\providecommand \BibitemShut  [1]{\csname bibitem#1\endcsname}%
\let\auto@bib@innerbib\@empty
\bibitem [{\citenamefont {Landau}(1957)}]{Landau1957}%
  \BibitemOpen
  \bibfield  {author} {\bibinfo {author} {\bibfnamefont {L.~D.}\ \bibnamefont
  {Landau}},\ }\href@noop {} {\bibfield  {journal} {\bibinfo  {journal} {J.
  Exp. Theor. Phys.}\ }\textbf {\bibinfo {volume} {5}},\ \bibinfo {pages} {101}
  (\bibinfo {year} {1957})}\BibitemShut {NoStop}%
\bibitem [{\citenamefont {Landau}(1959)}]{Landau1959}%
  \BibitemOpen
  \bibfield  {author} {\bibinfo {author} {\bibfnamefont {L.~D.}\ \bibnamefont
  {Landau}},\ }\href@noop {} {\bibfield  {journal} {\bibinfo  {journal} {J.
  Exp. Theor. Phys.}\ }\textbf {\bibinfo {volume} {8}},\ \bibinfo {pages} {70}
  (\bibinfo {year} {1959})}\BibitemShut {NoStop}%
\bibitem [{\citenamefont {Abrikosov}\ and\ \citenamefont
  {Khalatnikov}(1959)}]{Abrikosov1959}%
  \BibitemOpen
  \bibfield  {author} {\bibinfo {author} {\bibfnamefont {A.~A.}\ \bibnamefont
  {Abrikosov}}\ and\ \bibinfo {author} {\bibfnamefont {I.~M.}\ \bibnamefont
  {Khalatnikov}},\ }\href@noop {} {\bibfield  {journal} {\bibinfo  {journal}
  {Rep. Prog. Physics}\ }\textbf {\bibinfo {volume} {22}},\ \bibinfo {pages}
  {329} (\bibinfo {year} {1959})}\BibitemShut {NoStop}%
\bibitem [{\citenamefont {Hone}(1962)}]{Hone1962}%
  \BibitemOpen
  \bibfield  {author} {\bibinfo {author} {\bibfnamefont {D.}~\bibnamefont
  {Hone}},\ }\href {\doibase 10.1103/PhysRev.125.1494} {\bibfield  {journal}
  {\bibinfo  {journal} {Phys. Rev.}\ }\textbf {\bibinfo {volume} {125}},\
  \bibinfo {pages} {1494} (\bibinfo {year} {1962})}\BibitemShut {NoStop}%
\bibitem [{\citenamefont {Emery}\ and\ \citenamefont
  {Cheng}(1968)}]{Emery1968}%
  \BibitemOpen
  \bibfield  {author} {\bibinfo {author} {\bibfnamefont {V.~J.}\ \bibnamefont
  {Emery}}\ and\ \bibinfo {author} {\bibfnamefont {D.}~\bibnamefont {Cheng}},\
  }\href@noop {} {\bibfield  {journal} {\bibinfo  {journal} {Phys. Rev. Lett.}\
  }\textbf {\bibinfo {volume} {21}},\ \bibinfo {pages} {533} (\bibinfo {year}
  {1968})}\BibitemShut {NoStop}%
\bibitem [{\citenamefont {Rice}(1967{\natexlab{a}})}]{Rice1967a}%
  \BibitemOpen
  \bibfield  {author} {\bibinfo {author} {\bibfnamefont {M.~J.}\ \bibnamefont
  {Rice}},\ }\href@noop {} {\bibfield  {journal} {\bibinfo  {journal} {Phys.
  Rev.}\ }\textbf {\bibinfo {volume} {162}},\ \bibinfo {pages} {189} (\bibinfo
  {year} {1967}{\natexlab{a}})}\BibitemShut {NoStop}%
\bibitem [{\citenamefont {Rice}(1967{\natexlab{b}})}]{Rice1967b}%
  \BibitemOpen
  \bibfield  {author} {\bibinfo {author} {\bibfnamefont {M.~J.}\ \bibnamefont
  {Rice}},\ }\href@noop {} {\bibfield  {journal} {\bibinfo  {journal} {Phys.
  Rev.}\ }\textbf {\bibinfo {volume} {159}},\ \bibinfo {pages} {153} (\bibinfo
  {year} {1967}{\natexlab{b}})}\BibitemShut {NoStop}%
\bibitem [{\citenamefont {Pethick}(1969{\natexlab{a}})}]{Pethick1969-2}%
  \BibitemOpen
  \bibfield  {author} {\bibinfo {author} {\bibfnamefont {C.~J.}\ \bibnamefont
  {Pethick}},\ }\href {\doibase 10.1103/PhysRev.177.391} {\bibfield  {journal}
  {\bibinfo  {journal} {Phys. Rev.}\ }\textbf {\bibinfo {volume} {177}},\
  \bibinfo {pages} {391} (\bibinfo {year} {1969}{\natexlab{a}})}\BibitemShut
  {NoStop}%
\bibitem [{\citenamefont {Ron}(1970)}]{Ron1970}%
  \BibitemOpen
  \bibfield  {author} {\bibinfo {author} {\bibfnamefont {A.}~\bibnamefont
  {Ron}},\ }\href {\doibase 10.1103/PhysRevA.1.1211} {\bibfield  {journal}
  {\bibinfo  {journal} {Phys. Rev. A}\ }\textbf {\bibinfo {volume} {1}},\
  \bibinfo {pages} {1211} (\bibinfo {year} {1970})}\BibitemShut {NoStop}%
\bibitem [{\citenamefont {Abel}\ \emph {et~al.}(1966)\citenamefont {Abel},
  \citenamefont {Anderson},\ and\ \citenamefont {Wheatley}}]{Abel1966}%
  \BibitemOpen
  \bibfield  {author} {\bibinfo {author} {\bibfnamefont {W.~R.}\ \bibnamefont
  {Abel}}, \bibinfo {author} {\bibfnamefont {A.~C.}\ \bibnamefont {Anderson}},
  \ and\ \bibinfo {author} {\bibfnamefont {J.~C.}\ \bibnamefont {Wheatley}},\
  }\href@noop {} {\bibfield  {journal} {\bibinfo  {journal} {Phys. Rev. Lett.}\
  }\textbf {\bibinfo {volume} {17}},\ \bibinfo {pages} {74} (\bibinfo {year}
  {1966})}\BibitemShut {NoStop}%
\bibitem [{\citenamefont {Pethick}(1969{\natexlab{b}})}]{Pethick1969}%
  \BibitemOpen
  \bibfield  {author} {\bibinfo {author} {\bibfnamefont {C.~J.}\ \bibnamefont
  {Pethick}},\ }\href@noop {} {\bibfield  {journal} {\bibinfo  {journal} {Phy.
  Rev.}\ }\textbf {\bibinfo {volume} {185}},\ \bibinfo {pages} {384} (\bibinfo
  {year} {1969}{\natexlab{b}})}\BibitemShut {NoStop}%
\bibitem [{\citenamefont {Black}\ \emph {et~al.}(1971)\citenamefont {Black},
  \citenamefont {Hall},\ and\ \citenamefont {Thompson}}]{Black1971}%
  \BibitemOpen
  \bibfield  {author} {\bibinfo {author} {\bibfnamefont {M.~A.}\ \bibnamefont
  {Black}}, \bibinfo {author} {\bibfnamefont {H.~E.}\ \bibnamefont {Hall}}, \
  and\ \bibinfo {author} {\bibfnamefont {K.}~\bibnamefont {Thompson}},\
  }\href@noop {} {\bibfield  {journal} {\bibinfo  {journal} {J. Phys. C}\
  }\textbf {\bibinfo {volume} {4}},\ \bibinfo {pages} {129} (\bibinfo {year}
  {1971})}\BibitemShut {NoStop}%
\bibitem [{\citenamefont {Andronikashvili}(1946)}]{Andronikashvili1946}%
  \BibitemOpen
  \bibfield  {author} {\bibinfo {author} {\bibfnamefont {E.~L.}\ \bibnamefont
  {Andronikashvili}},\ }\href@noop {} {\bibfield  {journal} {\bibinfo
  {journal} {Zh. Eksp. Teor. Fiz.}\ }\textbf {\bibinfo {volume} {16}},\
  \bibinfo {pages} {780} (\bibinfo {year} {1946})}\BibitemShut {NoStop}%
\bibitem [{\citenamefont {Betts}\ \emph {et~al.}(1963)\citenamefont {Betts},
  \citenamefont {Osborne}, \citenamefont {Welber},\ and\ \citenamefont
  {Wilks}}]{Betts1963}%
  \BibitemOpen
  \bibfield  {author} {\bibinfo {author} {\bibfnamefont {D.~S.}\ \bibnamefont
  {Betts}}, \bibinfo {author} {\bibfnamefont {D.~W.}\ \bibnamefont {Osborne}},
  \bibinfo {author} {\bibfnamefont {B.}~\bibnamefont {Welber}}, \ and\ \bibinfo
  {author} {\bibfnamefont {J.}~\bibnamefont {Wilks}},\ }\href@noop {}
  {\bibfield  {journal} {\bibinfo  {journal} {Philos. Mag.}\ }\textbf {\bibinfo
  {volume} {8}},\ \bibinfo {pages} {977} (\bibinfo {year} {1963})}\BibitemShut
  {NoStop}%
\bibitem [{\citenamefont {Karrai}\ and\ \citenamefont
  {Grober}(1995)}]{Karrai1995}%
  \BibitemOpen
  \bibfield  {author} {\bibinfo {author} {\bibfnamefont {K.}~\bibnamefont
  {Karrai}}\ and\ \bibinfo {author} {\bibfnamefont {R.~D.}\ \bibnamefont
  {Grober}},\ }\href@noop {} {\bibfield  {journal} {\bibinfo  {journal}
  {Ultramicroscopy}\ }\textbf {\bibinfo {volume} {61}},\ \bibinfo {pages} {197}
  (\bibinfo {year} {1995})}\BibitemShut {NoStop}%
\bibitem [{\citenamefont {Blaauwgeers}\ \emph {et~al.}(2007)\citenamefont
  {Blaauwgeers}, \citenamefont {Blazkova}, \citenamefont {Clovecko},
  \citenamefont {Eltsov}, \citenamefont {de~Graaf}, \citenamefont {Hosio},
  \citenamefont {Krusius}, \citenamefont {Schmoranzer}, \citenamefont
  {Schoepe}, \citenamefont {Skrbek}, \citenamefont {Skyba}, \citenamefont
  {Solntsev},\ and\ \citenamefont {Zmeev}}]{Blaauwgeers2007}%
  \BibitemOpen
  \bibfield  {author} {\bibinfo {author} {\bibfnamefont {R.}~\bibnamefont
  {Blaauwgeers}}, \bibinfo {author} {\bibfnamefont {M.}~\bibnamefont
  {Blazkova}}, \bibinfo {author} {\bibfnamefont {M.}~\bibnamefont {Clovecko}},
  \bibinfo {author} {\bibfnamefont {V.~B.}\ \bibnamefont {Eltsov}}, \bibinfo
  {author} {\bibfnamefont {R.}~\bibnamefont {de~Graaf}}, \bibinfo {author}
  {\bibfnamefont {J.}~\bibnamefont {Hosio}}, \bibinfo {author} {\bibfnamefont
  {M.}~\bibnamefont {Krusius}}, \bibinfo {author} {\bibfnamefont
  {D.}~\bibnamefont {Schmoranzer}}, \bibinfo {author} {\bibfnamefont
  {W.}~\bibnamefont {Schoepe}}, \bibinfo {author} {\bibfnamefont
  {L.}~\bibnamefont {Skrbek}}, \bibinfo {author} {\bibfnamefont
  {P.}~\bibnamefont {Skyba}}, \bibinfo {author} {\bibfnamefont {R.~E.}\
  \bibnamefont {Solntsev}}, \ and\ \bibinfo {author} {\bibfnamefont {D.~E.}\
  \bibnamefont {Zmeev}},\ }\href@noop {} {\bibfield  {journal} {\bibinfo
  {journal} {J. Low Temp. Phys.}\ }\textbf {\bibinfo {volume} {146}},\ \bibinfo
  {pages} {537} (\bibinfo {year} {2007})}\BibitemShut {NoStop}%
\bibitem [{\citenamefont {Gonzalez}\ \emph {et~al.}(2013)\citenamefont
  {Gonzalez}, \citenamefont {Zheng}, \citenamefont {Garcell}, \citenamefont
  {Lee},\ and\ \citenamefont {Chan}}]{Gonzalez2013}%
  \BibitemOpen
  \bibfield  {author} {\bibinfo {author} {\bibfnamefont {M.}~\bibnamefont
  {Gonzalez}}, \bibinfo {author} {\bibfnamefont {P.}~\bibnamefont {Zheng}},
  \bibinfo {author} {\bibfnamefont {E.}~\bibnamefont {Garcell}}, \bibinfo
  {author} {\bibfnamefont {Y.}~\bibnamefont {Lee}}, \ and\ \bibinfo {author}
  {\bibfnamefont {H.~B.}\ \bibnamefont {Chan}},\ }\href@noop {} {\bibfield
  {journal} {\bibinfo  {journal} {Rev. Sci. Instrum.}\ }\textbf {\bibinfo
  {volume} {84}},\ \bibinfo {pages} {025003} (\bibinfo {year}
  {2013})}\BibitemShut {NoStop}%
\bibitem [{\citenamefont {Collin}\ \emph {et~al.}(2010)\citenamefont {Collin},
  \citenamefont {Kofler}, \citenamefont {Heron}, \citenamefont {Bourgeois},
  \citenamefont {Bunkov},\ and\ \citenamefont {Godfrin}}]{Collin2010}%
  \BibitemOpen
  \bibfield  {author} {\bibinfo {author} {\bibfnamefont {E.}~\bibnamefont
  {Collin}}, \bibinfo {author} {\bibfnamefont {J.}~\bibnamefont {Kofler}},
  \bibinfo {author} {\bibfnamefont {J.-S.}\ \bibnamefont {Heron}}, \bibinfo
  {author} {\bibfnamefont {O.}~\bibnamefont {Bourgeois}}, \bibinfo {author}
  {\bibfnamefont {Y.~M.}\ \bibnamefont {Bunkov}}, \ and\ \bibinfo {author}
  {\bibfnamefont {H.}~\bibnamefont {Godfrin}},\ }\href@noop {} {\bibfield
  {journal} {\bibinfo  {journal} {J Low Temp Phys}\ }\textbf {\bibinfo {volume}
  {158}},\ \bibinfo {pages} {678} (\bibinfo {year} {2010})}\BibitemShut
  {NoStop}%
\bibitem [{\citenamefont {Collin}\ \emph {et~al.}(2011)\citenamefont {Collin},
  \citenamefont {Moutonet}, \citenamefont {Heron}, \citenamefont {Bourgeois},
  \citenamefont {Bunkov},\ and\ \citenamefont {Godfrin}}]{Collin2011}%
  \BibitemOpen
  \bibfield  {author} {\bibinfo {author} {\bibfnamefont {E.}~\bibnamefont
  {Collin}}, \bibinfo {author} {\bibfnamefont {T.}~\bibnamefont {Moutonet}},
  \bibinfo {author} {\bibfnamefont {J.-S.}\ \bibnamefont {Heron}}, \bibinfo
  {author} {\bibfnamefont {O.}~\bibnamefont {Bourgeois}}, \bibinfo {author}
  {\bibfnamefont {Y.~M.}\ \bibnamefont {Bunkov}}, \ and\ \bibinfo {author}
  {\bibfnamefont {H.}~\bibnamefont {Godfrin}},\ }\href@noop {} {\bibfield
  {journal} {\bibinfo  {journal} {J Low Temp Phys}\ }\textbf {\bibinfo {volume}
  {162}},\ \bibinfo {pages} {653} (\bibinfo {year} {2011})}\BibitemShut
  {NoStop}%
\bibitem [{\citenamefont {Rojas}\ and\ \citenamefont
  {Davis}(2015)}]{Rojas2015}%
  \BibitemOpen
  \bibfield  {author} {\bibinfo {author} {\bibfnamefont {X.}~\bibnamefont
  {Rojas}}\ and\ \bibinfo {author} {\bibfnamefont {J.~P.}\ \bibnamefont
  {Davis}},\ }\href {\doibase 10.1103/PhysRevB.91.024503} {\bibfield  {journal}
  {\bibinfo  {journal} {Phys. Rev. B}\ }\textbf {\bibinfo {volume} {91}},\
  \bibinfo {pages} {024503} (\bibinfo {year} {2015})}\BibitemShut {NoStop}%
\bibitem [{\citenamefont {Vorontsov}\ and\ \citenamefont
  {Sauls}(2007)}]{Vorontsov2007}%
  \BibitemOpen
  \bibfield  {author} {\bibinfo {author} {\bibfnamefont {A.~B.}\ \bibnamefont
  {Vorontsov}}\ and\ \bibinfo {author} {\bibfnamefont {J.~A.}\ \bibnamefont
  {Sauls}},\ }\href {\doibase 10.1103/PhysRevLett.98.045301} {\bibfield
  {journal} {\bibinfo  {journal} {Phys. Rev. Lett.}\ }\textbf {\bibinfo
  {volume} {98}},\ \bibinfo {pages} {045301} (\bibinfo {year}
  {2007})}\BibitemShut {NoStop}%
\bibitem [{\citenamefont {Sharma}\ \emph {et~al.}(2011)\citenamefont {Sharma},
  \citenamefont {Corcoles}, \citenamefont {Bennett}, \citenamefont {Parpia},
  \citenamefont {Cowan}, \citenamefont {Casey},\ and\ \citenamefont
  {Saunders}}]{Sharma2011}%
  \BibitemOpen
  \bibfield  {author} {\bibinfo {author} {\bibfnamefont {P.}~\bibnamefont
  {Sharma}}, \bibinfo {author} {\bibfnamefont {A.}~\bibnamefont {Corcoles}},
  \bibinfo {author} {\bibfnamefont {R.~G.}\ \bibnamefont {Bennett}}, \bibinfo
  {author} {\bibfnamefont {J.~M.}\ \bibnamefont {Parpia}}, \bibinfo {author}
  {\bibfnamefont {B.}~\bibnamefont {Cowan}}, \bibinfo {author} {\bibfnamefont
  {A.}~\bibnamefont {Casey}}, \ and\ \bibinfo {author} {\bibfnamefont
  {J.}~\bibnamefont {Saunders}},\ }\href@noop {} {\bibfield  {journal}
  {\bibinfo  {journal} {Phys. Rev. Lett.}\ }\textbf {\bibinfo {volume} {107}},\
  \bibinfo {pages} {196805} (\bibinfo {year} {2011})}\BibitemShut {NoStop}%
\bibitem [{\citenamefont {Barquist}\ \emph {et~al.}(2014)\citenamefont
  {Barquist}, \citenamefont {Bauer}, \citenamefont {Edmunds}, \citenamefont
  {Zheng}, \citenamefont {Jiang}, \citenamefont {Gonzalez}, \citenamefont
  {Lee},\ and\ \citenamefont {Chan}}]{Barquist2014}%
  \BibitemOpen
  \bibfield  {author} {\bibinfo {author} {\bibfnamefont {C.~S.}\ \bibnamefont
  {Barquist}}, \bibinfo {author} {\bibfnamefont {J.}~\bibnamefont {Bauer}},
  \bibinfo {author} {\bibfnamefont {T.}~\bibnamefont {Edmunds}}, \bibinfo
  {author} {\bibfnamefont {P.}~\bibnamefont {Zheng}}, \bibinfo {author}
  {\bibfnamefont {W.~G.}\ \bibnamefont {Jiang}}, \bibinfo {author}
  {\bibfnamefont {M.}~\bibnamefont {Gonzalez}}, \bibinfo {author}
  {\bibfnamefont {Y.}~\bibnamefont {Lee}}, \ and\ \bibinfo {author}
  {\bibfnamefont {H.~B.}\ \bibnamefont {Chan}},\ }\href@noop {} {\bibfield
  {journal} {\bibinfo  {journal} {J. Phys.: Conf. Ser.}\ }\textbf {\bibinfo
  {volume} {568}},\ \bibinfo {pages} {032003} (\bibinfo {year}
  {2014})}\BibitemShut {NoStop}%
\bibitem [{\citenamefont {Landau}\ and\ \citenamefont
  {Lifshitz}(2008)}]{LandauFM}%
  \BibitemOpen
  \bibfield  {author} {\bibinfo {author} {\bibfnamefont {L.~D.}\ \bibnamefont
  {Landau}}\ and\ \bibinfo {author} {\bibfnamefont {E.~M.}\ \bibnamefont
  {Lifshitz}},\ }\href@noop {} {\emph {\bibinfo {title} {Fluid Mechanics}}},\
  \bibinfo {edition} {2nd}\ ed.,\ edited by\ \bibinfo {editor} {\bibfnamefont
  {J.~B.}\ \bibnamefont {Sykes}}\ and\ \bibinfo {editor} {\bibfnamefont
  {W.~H.}\ \bibnamefont {Reid}},\ \bibinfo {series} {Course of Theoretical
  Physics}, Vol.~\bibinfo {volume} {6}\ (\bibinfo  {publisher} {Elsevier
  Science Publisher},\ \bibinfo {year} {2008})\BibitemShut {NoStop}%
\bibitem [{\citenamefont {Veijola}\ and\ \citenamefont
  {Turowski}(2001)}]{Veijola2001}%
  \BibitemOpen
  \bibfield  {author} {\bibinfo {author} {\bibfnamefont {T.}~\bibnamefont
  {Veijola}}\ and\ \bibinfo {author} {\bibfnamefont {M.}~\bibnamefont
  {Turowski}},\ }\href@noop {} {\bibfield  {journal} {\bibinfo  {journal} {J.
  Microelectromech. Syst.}\ }\textbf {\bibinfo {volume} {10}},\ \bibinfo
  {pages} {263} (\bibinfo {year} {2001})}\BibitemShut {NoStop}%
\bibitem [{\citenamefont {Einzel}\ \emph {et~al.}(1990)\citenamefont {Einzel},
  \citenamefont {Panzer},\ and\ \citenamefont {Liu}}]{Einzel1990}%
  \BibitemOpen
  \bibfield  {author} {\bibinfo {author} {\bibfnamefont {D.}~\bibnamefont
  {Einzel}}, \bibinfo {author} {\bibfnamefont {P.}~\bibnamefont {Panzer}}, \
  and\ \bibinfo {author} {\bibfnamefont {M.}~\bibnamefont {Liu}},\ }\href@noop
  {} {\bibfield  {journal} {\bibinfo  {journal} {Phys. Rev. Lett.}\ }\textbf
  {\bibinfo {volume} {64}},\ \bibinfo {pages} {2269} (\bibinfo {year}
  {1990})}\BibitemShut {NoStop}%
\bibitem [{\citenamefont {Einzel}\ and\ \citenamefont
  {Parpia}(1997)}]{Einzel1997}%
  \BibitemOpen
  \bibfield  {author} {\bibinfo {author} {\bibfnamefont {D.}~\bibnamefont
  {Einzel}}\ and\ \bibinfo {author} {\bibfnamefont {J.~M.}\ \bibnamefont
  {Parpia}},\ }\href@noop {} {\bibfield  {journal} {\bibinfo  {journal} {J. Low
  Temp. Phys.}\ }\textbf {\bibinfo {volume} {109}},\ \bibinfo {pages} {1}
  (\bibinfo {year} {1997})}\BibitemShut {NoStop}%
\bibitem [{\citenamefont {Halperin}\ and\ \citenamefont
  {Pitaevskii}(1990)}]{Halperin1990}%
  \BibitemOpen
  \bibfield  {author} {\bibinfo {author} {\bibfnamefont {W.~P.}\ \bibnamefont
  {Halperin}}\ and\ \bibinfo {author} {\bibfnamefont {L.~P.}\ \bibnamefont
  {Pitaevskii}},\ }\href@noop {} {\emph {\bibinfo {title} {Helium Three}}},\
  edited by\ \bibinfo {editor} {\bibfnamefont {W.~P.}\ \bibnamefont
  {Halperin}}\ and\ \bibinfo {editor} {\bibfnamefont {L.~P.}\ \bibnamefont
  {Pitaevskii}}\ (\bibinfo  {publisher} {Elsevier Science Publisher},\ \bibinfo
  {year} {1990})\BibitemShut {NoStop}%
\bibitem [{\citenamefont {Wheatley}(1975)}]{Wheatley1975}%
  \BibitemOpen
  \bibfield  {author} {\bibinfo {author} {\bibfnamefont {J.~C.}\ \bibnamefont
  {Wheatley}},\ }\href@noop {} {\bibfield  {journal} {\bibinfo  {journal} {Rev.
  Mod. Phys.}\ }\textbf {\bibinfo {volume} {47}},\ \bibinfo {pages} {415}
  (\bibinfo {year} {1975})}\BibitemShut {NoStop}%
\bibitem [{\citenamefont {Ketterson}\ \emph {et~al.}(1975)\citenamefont
  {Ketterson}, \citenamefont {Roach}, \citenamefont {Abraham},\ and\
  \citenamefont {Roach}}]{Ketterson1975}%
  \BibitemOpen
  \bibfield  {author} {\bibinfo {author} {\bibfnamefont {J.~B.}\ \bibnamefont
  {Ketterson}}, \bibinfo {author} {\bibfnamefont {P.~R.}\ \bibnamefont
  {Roach}}, \bibinfo {author} {\bibfnamefont {B.~M.}\ \bibnamefont {Abraham}},
  \ and\ \bibinfo {author} {\bibfnamefont {P.~D.}\ \bibnamefont {Roach}},\
  }\enquote {\bibinfo {title} {Quantum statistics and the many-body problem},}\
  \ (\bibinfo  {publisher} {Plenum, New York},\ \bibinfo {year} {1975})\
  p.~\bibinfo {pages} {35}\BibitemShut {NoStop}%
\bibitem [{\citenamefont {Savard}\ \emph {et~al.}(2009)\citenamefont {Savard},
  \citenamefont {Tremblay-Darveau},\ and\ \citenamefont
  {Gervais}}]{Savard2009}%
  \BibitemOpen
  \bibfield  {author} {\bibinfo {author} {\bibfnamefont {M.}~\bibnamefont
  {Savard}}, \bibinfo {author} {\bibfnamefont {C.}~\bibnamefont
  {Tremblay-Darveau}}, \ and\ \bibinfo {author} {\bibfnamefont
  {G.}~\bibnamefont {Gervais}},\ }\href@noop {} {\bibfield  {journal} {\bibinfo
   {journal} {Phys. Rev. Lett.}\ }\textbf {\bibinfo {volume} {103}},\ \bibinfo
  {pages} {104502} (\bibinfo {year} {2009})}\BibitemShut {NoStop}%
\bibitem [{\citenamefont {Takeuchi}\ \emph {et~al.}(2012)\citenamefont
  {Takeuchi}, \citenamefont {Higashitani}, \citenamefont {Nagai}, \citenamefont
  {Choi}, \citenamefont {Moon}, \citenamefont {Masuhara}, \citenamefont
  {Meisel}, \citenamefont {Lee},\ and\ \citenamefont {Mulders}}]{Takeuchi2012}%
  \BibitemOpen
  \bibfield  {author} {\bibinfo {author} {\bibfnamefont {H.}~\bibnamefont
  {Takeuchi}}, \bibinfo {author} {\bibfnamefont {S.}~\bibnamefont
  {Higashitani}}, \bibinfo {author} {\bibfnamefont {K.}~\bibnamefont {Nagai}},
  \bibinfo {author} {\bibfnamefont {H.~C.}\ \bibnamefont {Choi}}, \bibinfo
  {author} {\bibfnamefont {B.~H.}\ \bibnamefont {Moon}}, \bibinfo {author}
  {\bibfnamefont {N.}~\bibnamefont {Masuhara}}, \bibinfo {author}
  {\bibfnamefont {M.~W.}\ \bibnamefont {Meisel}}, \bibinfo {author}
  {\bibfnamefont {Y.}~\bibnamefont {Lee}}, \ and\ \bibinfo {author}
  {\bibfnamefont {N.}~\bibnamefont {Mulders}},\ }\href@noop {} {\bibfield
  {journal} {\bibinfo  {journal} {Phys. Rev. Lett.}\ }\textbf {\bibinfo
  {volume} {108}},\ \bibinfo {pages} {225307} (\bibinfo {year}
  {2012})}\BibitemShut {NoStop}%
\bibitem [{\citenamefont {Einzel}\ and\ \citenamefont
  {Parpia}(1998)}]{Einzel1998}%
  \BibitemOpen
  \bibfield  {author} {\bibinfo {author} {\bibfnamefont {D.}~\bibnamefont
  {Einzel}}\ and\ \bibinfo {author} {\bibfnamefont {J.~M.}\ \bibnamefont
  {Parpia}},\ }\href@noop {} {\bibfield  {journal} {\bibinfo  {journal} {Phys.
  Rev. Lett.}\ }\textbf {\bibinfo {volume} {81}},\ \bibinfo {pages} {3896}
  (\bibinfo {year} {1998})}\BibitemShut {NoStop}%
\end{thebibliography}
%

\end{document}